\documentclass[preprint,epsf,psfig]{revtex4}
\usepackage{amssymb}
\usepackage{epsfig}
\DeclareMathAlphabet{\pazocal}{OMS}{zplm}{m}{n}            
\usepackage{graphicx}
\usepackage{color}
\usepackage{bm}
\usepackage[normalem]{ulem}     
\usepackage{soul}
\usepackage{amsmath}
\usepackage{braket} 
\usepackage{rotating}
\usepackage{multirow} 
\DeclareMathAlphabet{\pazocal}{OMS}{zplm}{m}{n}            
\usepackage{graphicx}
\usepackage{hyperref}

\begin{document}


\title{Polar metals: Principles and Prospects}

\author{Sayantika Bhowal} 
\affiliation{Materials Theory, ETH Zurich, Wolfgang-Pauli-Strasse 27, 8093 Zurich, Switzerland} 

\author{Nicola A. Spaldin}
\affiliation{Materials Theory, ETH Zurich, Wolfgang-Pauli-Strasse 27, 8093 Zurich, Switzerland}

\begin{abstract}
We review the class of materials known as polar metals, in which polarity and metallicity coexist in the same phase. While the notion of polar metals was first invoked more than 50 years ago, their practical realization has proved challenging, since the itinerant carriers required for metallicity tend to screen any polarization. Huge progress has been made in the last decade, with many mechanisms for combining polarity and metallicity proposed, and the first examples, LiOsO$_3$ and WTe$_2$, identified experimentally. The availability of polar metallic samples has opened a new paradigm in polar metal research, with implications in the fields of topology, ferroelectricity, magnetoelectricity, spintronics, and superconductivity. Here, we review the principles and techniques that have been developed to design and engineer polar metals and describe some of their interesting properties, with a focus on the most promising directions for future work.
\end{abstract}

\maketitle


\section{INTRODUCTION}

Combining multiple physical phenomena into a single system is an important route to designing materials with improved and innovative properties. Of particular interest is the coexistence of behaviors  that are usually considered to be mutually exclusive. Examples include the coexistence of ferroelectricity and magnetism, where the former is favored in empty $d$-shell systems while the latter occurs in systems with partially filled $d$ orbitals, ferromagnetism and superconductivity, in which the magnetism is expected to destroy the pair interaction responsible for superconductivity, and 
high electrical conductivity combined with low thermal conductivity, which is unexpected since charge-carrying electrons also transport heat. In addition to the intriguing fundamental physics and challenging chemistry, such combinations are promising for new technologies based on electric-field-control of magnetism, vortex manipulation of spin textures, or thermoelectric conversion. 
In this review, we focus an exciting new class of materials with properties that tend not to coexist, the family of {\it polar metals}. We begin by discussing the origins of the contra-indication between polarization and metallicity. We then discuss strategies for engineering polar metals, based both on introducing metallicity into known polar systems and introducing polarity into known metallic systems, as well as through entirely novel mechanisms. The final part of the review describes the exotic physics that has been discovered or predicted in polar metallic systems, followed by our thoughts for promising future directions.

\subsection{Some Definitions: ``Polar'', ``Ferroelectric'' And ``Metal''} \label{PolarMetal}

While the meaning of the term ``polar metal'' is seemingly obvious, it has been used in a variety of contexts to imply rather different meanings. Therefore, we begin by reviewing the formal definitions of both {\it metal} and {\it polar}, as well as the related {\it ferroelectric}, which will also be important for our discussion. 

\subsubsection{Polar} 
\begin{figure*}[h]
\includegraphics[width=\columnwidth]{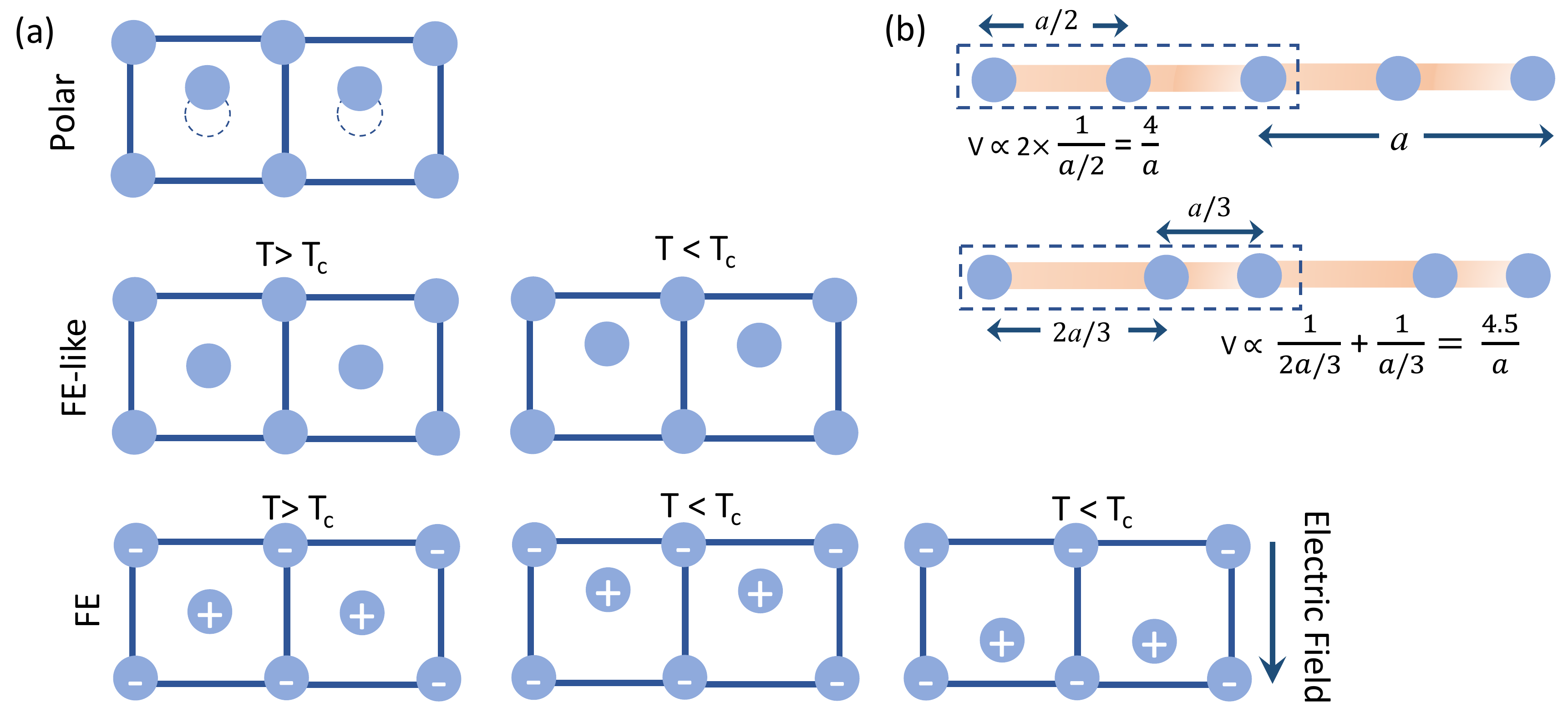}
\caption{Introduction to polar metals. (a) Schematic illustration of polar metal ({\it top}), FE-like metal ({\it middle}), and FE-metal ({\it bottom}). (b) Cartoon showing 1D chains of metal ions, arranged on a centrosymmetric ({\it top}) and a noncentrosymmetric ({\it bottom}) lattice with smaller Coulomb repulsion energy V for the centrosymmetric lattice.
}
\label{fig1}
\end{figure*}

A {\it polar direction} is one for which the two directional senses are different. There are many possible polar directions in noncentrosymmetric crystal structures, since all directions that are neither normal to an even-fold rotation axis nor along a roto-inversion axis are polar. A {\it polar point group}, which is also referred to as a {\it polar crystal class}, on the other hand, is more restrictive, since it must contain a polar direction that allows (at least in principle) the existence of a permanent electric dipole moment; this can occur only along those polar directions that have no symmetry equivalent directions (see for example \textbf{Figure \ref{fig1}}\textit{a} upper panel). Such electric-dipole-moment-compatible polar directions occur along the rotation axis in the $6mm$, $6$, $4mm$, $4$, $3m$, $3$, $mm2$ and $2$ classes, parallel to the mirror plane in $m$, and along any direction in class $1$. Note that, while all of the polar crystal classes are noncentrosymmetric, that is they lack inversion symmetry, it is not the case that all noncentrosymmetric crystal structures are polar. 
Since the definition of a polar crystal class is based only on the crystallographic symmetry, any material belonging to one of these polar crystal classes is in this sense polar. It is not a requirement that any particular property associated with the existence of the permanent electric dipole moment should be measurable, and the definition is equally applicable to metallic or insulating systems. 
In contrast, the macroscopic {\it electric polarization} of a bulk sample is defined by the modern theory of polarization \cite{Resta/Vanderbilt:2007} to be the amount of charge that flows as the system is deformed from a non-polar reference structure to the polar state; the middle row of \textbf{Figure \ref{fig1}}\textit{a} shows such an example. The definition only holds for insulating systems, and so a metal can not have a macroscopic polarization (we will formalize this in the next section). Therefore, while the term {\it polar metal} can be used for metallic systems that have a polar crystal class, it should not be interpreted to imply the existence of an electric polarization. 

\subsubsection{Metal} The colloquial distinction between metals and insulators, as materials that do or do not conduct electricity respectively, is captured well by the classical picture of bonding in solids, in which the electrons in insulators are bound or localized to their parent ions, whereas those in metals are free. A quantum-mechanical extension is non-trivial, however, since quantum mechanically all electrons in a periodic solid are delocalized in Bloch states, whereas of course not all periodic solids are metallic. This is usually addressed in introductory classes using the concepts of band theory, with insulators and metals distinguished by the position of their Fermi energies -- within the band gap for insulators or within an energy band for metals. This simple picture, however, is only suitable for simple periodic band insulators, and fails to capture insulating behavior driven by electron correlations (so-called Mott insulators) or disorder (through Anderson localization). A rigorous distinction requires the theory of electron localization first introduced by Kohn \cite{Kohn:1964}, and subsequently formalized by Resta and Sorella \cite{Resta:1998,Resta/Sorella:1999}. This theory shows that the same many-body expectation value that discriminates between localized/insulating and delocalized/metallic ground states, also determines the macroscopic polarization, and it is therefore impossible for a metallic material to sustain an electric polarization. (For an excellent review see Reference~\citenum{Resta_2002}).  

\subsubsection{Ferroelectric} For completeness, we remind the reader that a ferroelectric (FE) material is one which has a spontaneous electric polarization that is switchable by an applied electric field (\textbf{Figure \ref{fig1}}\textit{a} lower panel). Insulating behavior is necessary both for the existence of the electric polarization and for the practical requirement of switching in an applied electric field. While it is not part of the definition, ferroelectricity is often associated with a structural phase transition from an inversion-symmetric non-polar paraelectric (PE) structure to a noncentrosymmetric polar structure as the temperature is reduced (\textbf{Figure \ref{fig1}}\textit{a} middle panel). In a typical FE such as PbTiO$_3$ (PTO), the phase transition is characterized by the softening of a transverse optical (TO) phonon \cite{Shirane1970}, with the frequency of the TO phonon decreasing with decreasing temperature and becoming zero at the transition temperature ($T_c$), when the ions off-center collectively from their centrosymmetric positions. Such a symmetry-lowering phase transition from a non-polar to a polar phase, accompanied by the softening of a TO phonon, can occur in a metal, with such materials referred to as {\it ferroelectric-like} polar metals, with the ``like'' indicating the existence of the phase transition but not implying switchability.

\subsection{Polar Metals} Since the coexistence of electric polarization and metallicity is formally contra-indicated, for the purposes of this review we define a polar metal to be a material that combines a polar crystal class with the existence of charge carriers that lead to a reasonable electronic conductivity. 
In addition to metallic conductivity, we include semiconducting, semi-metallic and polaronic conduction, but exclude cases in which any carriers are bound to defects at all reasonable temperatures. We do not require that the polar direction can be switched, even in principle, by an applied electric field.

As discussed above, there is no obvious contra-indication between metallicity and polar crystal classes, and so it is perhaps surprising that polar metals are so rare! 
Next, we address the questions of why most conventional metals are in fact 
centrosymmetric, and most polar, and in particular FE or FE-like materials are insulators.

\subsubsection{Why Are Conventional Metals Centrosymmetric?} \label{CentroInsulator}
Conventional metallic materials, such as copper or iron, adopt centrosymmetric, cubic or hexagonal structures. This is a consequence of their non-directional, metallic bonding and the Coulomb repulsion between the ions, which is minimized at a particular density for evenly spaced ions (see the cartoon of \textbf{Figure \ref{fig1}}\textit{b}). 
While an ordered metal alloy, for example of three different elements layered in an A B C configuration, would be polar, achieving such an ordered arrangement in a metallically bonded system is challenging. Metallic bonding in a heteroatomic system requires that the constituents have similar electronegativity, resulting in a low driving force for ordering that is easily outweighed by the entropic stabilization associated with disorder. 

\subsubsection{Why Are Conventional Ferroelectric-like Materials Insulating?}

The softening of the TO phonon and resulting phase transition from a centrosymmetric to a polar state discussed above occurs almost exclusively in insulating systems. This can be understood from two perspectives: First, in the traditional ``physics-based'' model picture, whether or not a FE-like instability prevails at low temperature is determined by a competition between two energies. These are the short-range Coulomb repulsions between the valence electron clouds around the ions, which favor the non-polar PE phase, and the long-range dipole-dipole Coulomb interactions, which favor the FE phase \cite{cohen1992}. The dipole-dipole interactions are screened by free carriers, so metallicity disfavors a FE-like distortion. The second-order Jahn-Teller effect provides a more local ``chemistry-based'' picture \cite{Burdett:1981}, showing that chemical bonding between cations and ligands, which leads in turn to  off-centering of the ions, is favored by empty-valence-shell cations. The prototype example is the formally empty $3d$ shell of the Ti$^{4+}$ ion in FE PTO or BaTiO$_3$ (BTO), in which the $d^0$ configuration correlates with insulating behavior.

\begin{figure*}[t]
\includegraphics[width=\columnwidth]{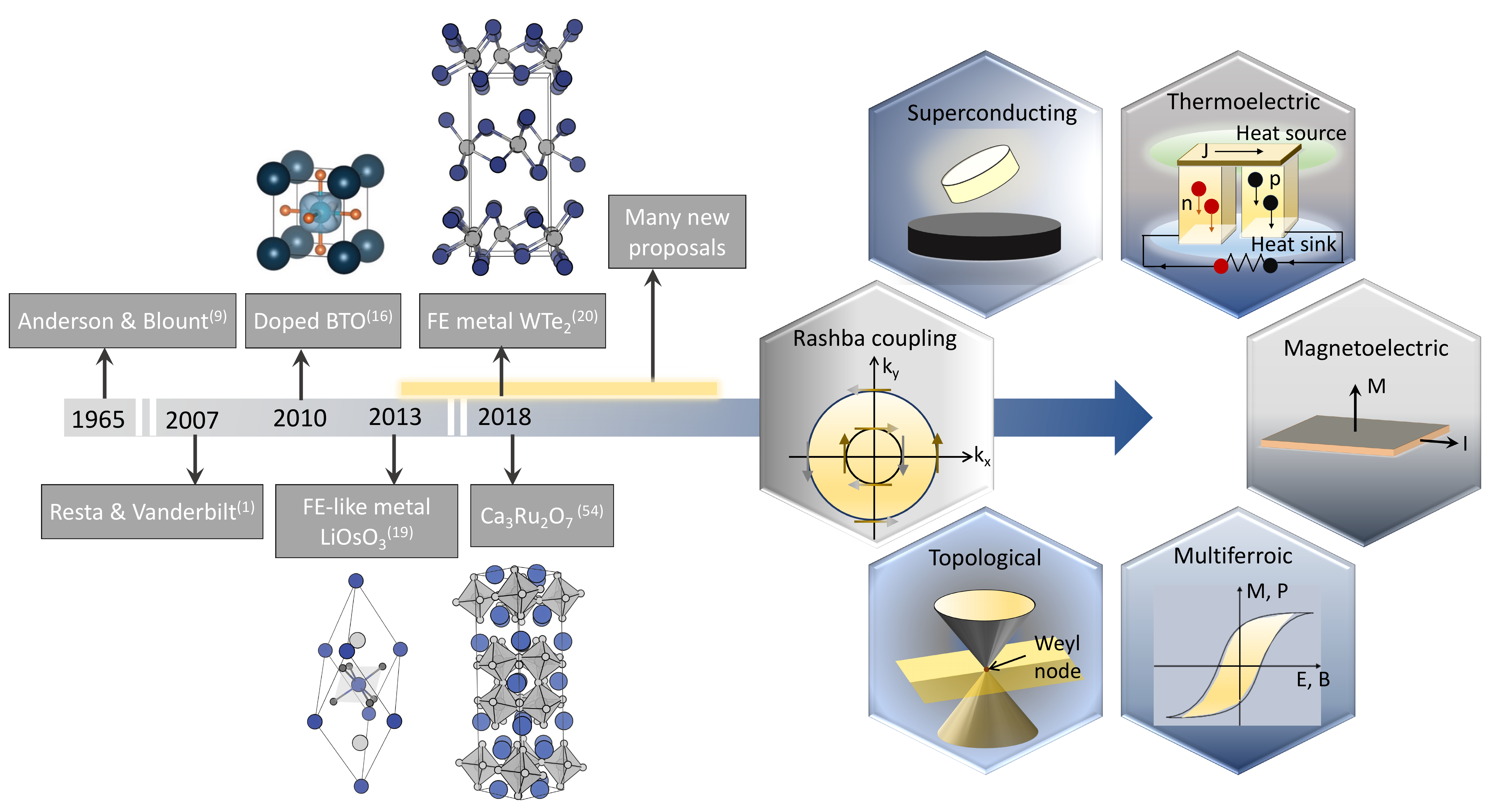}
\caption{Schematic showing the timeline of research (\textit{left}) on polar metals  and their exotic properties (\textit{right}).}
\label{fig0}
\end{figure*}

FE metals were first discussed, to our knowledge, in 1965 by Anderson and Blount \cite{AndersonBlount1965}, who suggested that a FE-like phase transition could occur in a metal if the metallic electrons do not interact strongly with the FE TO phonons. They proposed the known phase transition in V$_3$Si \cite{BattermanBarrett1964} as a candidate, although it was later established that strain rather than the polar distortion is the order parameter in this case \cite{Testardi1967,Brown_2001,Paduani2008}. Research languished for almost four decades, until the proposal of Cd$_2$Re$_2$O$_7$ as a promising candidate \cite{Sergienko2004}. While this was later characterized as a {\it piezoelectric} (i.e. noncentrosymmetric but non-polar) metal \cite{Ishibashi2010}, it revitalized activity in the field
and a number of other mechanisms and candidate materials were suggested \cite{Kolodiazhnyi2010,Filippetti2016,Urru2020}. The field was transformed by two milestones that occurred in the last decade,  the first unambiguous experimental realization \cite{Shi2013} of a FE-like metal with a symmetry-lowering phase transition, LiOsO$_3$, in 2013 and the first FE metal with a switchable polar axis, WTe$_2$, in 2018 \cite{Fei2018,Sharma2019}. The existence of real polar metallic materials has enabled in turn the measurement of novel properties (see \textbf{Figure \ref{fig0}}), resulting in an explosion of activity. 
Current work focuses on the discovery of new polar metals, motivated by exotic behaviors such as combined non-trivial topology and antisymmetric spin–orbit interactions, topologically protected spin currents and unconventional superconductivity, as well as their device potential as electrodes for stabilizing out-of-plane polarization in ultra-thin FE capacitors \cite{Puggioni2018}. 

Our goals in this article are two-fold. First, to present the reader with a classification of the different mechanisms and design principles, including material examples, that have led to or are promising for polar, FE-like and FE metals. Second, to discuss  the intriguing known and proposed properties of polar metals and their potential applicability. Throughout, we mention our favorite emerging directions and open questions, and emphasize the areas we feel should be addressed to accelerate future progress in the field. We note that a number of excellent reviews on polar metals already exist, and we point the reader in particular to Reference \citenum{Zhou2020} for a comprehensive discussion. 

\section{DESIGN PRINCIPLES AND MATERIALS}

Extensive research over the last decade has provided numerous candidate polar metals, as well as design strategies that serve as a guideline for future predictions \cite{Kim2016}. Here, we review some selected material types, which we believe have contributed significantly to the advancement of the field, as well as the design techniques that we find most promising for future investigation. We begin by analyzing examples in which carriers are added directly to conventional FEs, where they perturb the chemical bonds that are strongly involved in the soft TO phonon that is responsible for the ferroelectricity. While these violate the general Anderson and Blount principle \cite{AndersonBlount1965} of decoupling (or weakly coupling) the metallicity, characterized by the electrons at the Fermi level, and the ferroelectricity, driven by the soft TO phonons, we see that polar metallicity is possible in some cases. We then turn to unconventional FEs, with alternative driving forces for the TO phonon softening, and find that, as expected, the coexistence of metallicity and ferroelectricity is more robust. This is particularly the case in layered materials, for which two-dimensional metallicity coexists happily with perpendicular polarization in a capacitor geometry. Finally, we describe in some detail the two most promising polar metals identified to date, FE-like LiOsO$_3$, and layered WTe$_2$, for which polarization switching has been demonstrated.

\subsection{Adding Carriers To Polar Insulators And Conventional Ferroelectrics}

\begin{figure*}[t]
\includegraphics[scale=.28]{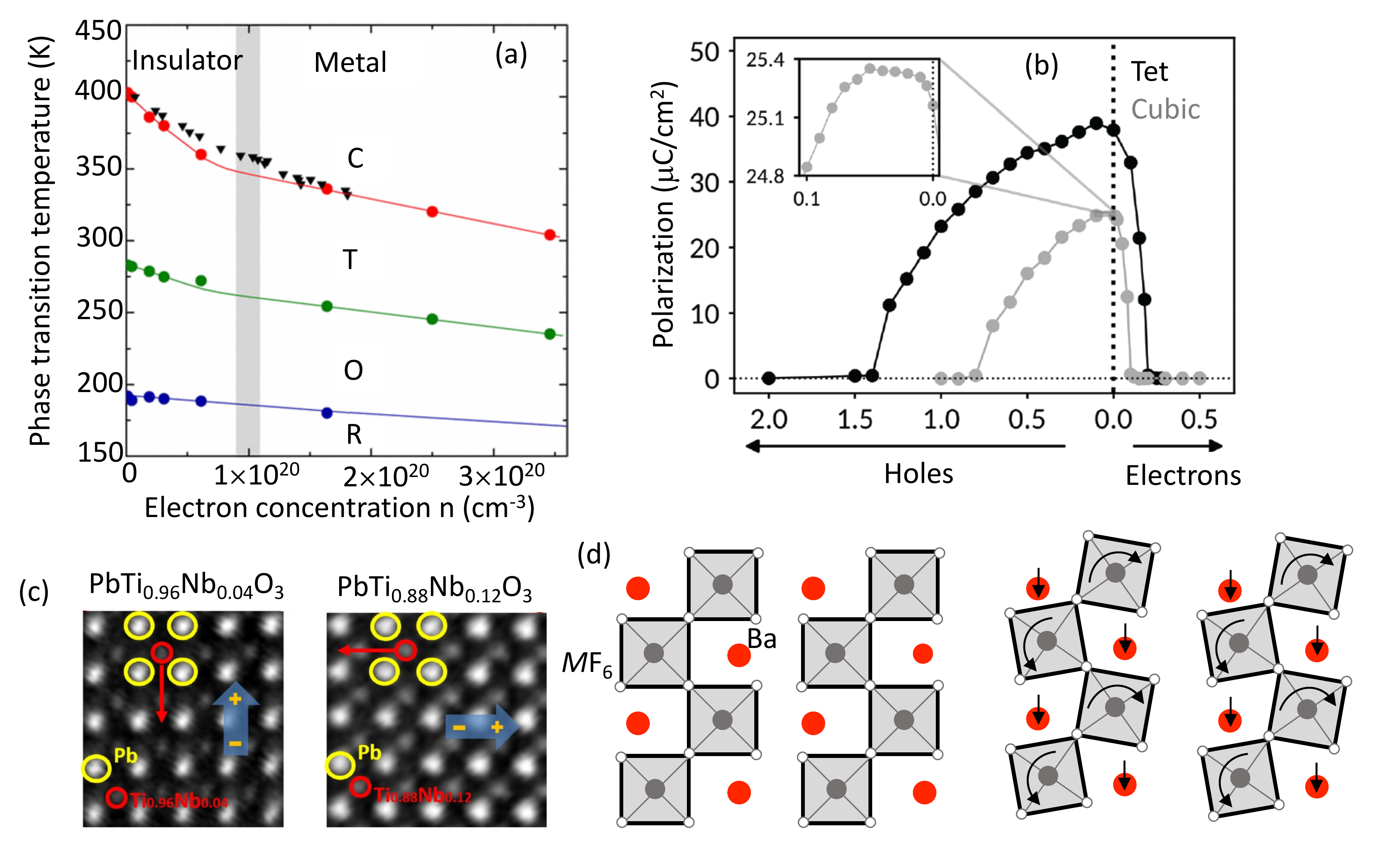}
\caption{Polar metals from doping FEs. (a) Temperature-dependent phase diagram of BaTiO$_{3-\delta}$ as a function of electron dopant concentration, $n$, indicating the persistence of structural phase transitions through the metal-insulator transition. C, T, O, and R represent the cubic, tetragonal, orthorhombic, and rhombohedral structural phases. (b) Polarization as a function of electron and hole doping calculated using density functional theory (DFT) for BTO with tetragonal (Tet) and pseudocubic (Cubic) lattice constants. 
For the same carrier concentration, the pseudocubic system has lower polarization than the tetragonal system, although the trends are similar for both cases. Inset shows the increase in polarization at small hole concentration. 
(c) STEM images of Nb-doped PTO for two different doping concentrations, showing the displacements (indicated by red arrows) of the Ti (Nb) atoms with respect to the four nearest Pb atoms inducing a local polarization (indicated by large blue arrows). (d) Structure of the layered proper geometric FE Ba$M$F$_4$, showing the high-symmetry inversion-symmetric structure (\textit{left}) and the polar structure (\textit{right}). The rotation of the $M$F$_6$ octahedra (indicated in curved arrows) forces the polar displacement of the Ba ions (indicated in straight arrows).   
 Panel (a) is reprinted with permission from Reference \citenum{Kolodiazhnyi2010} Copyright 2010 by the
American Physical Society, panel (b) is reproduced from Reference \citenum{Michel2021} with permission from the Royal Society of Chemistry, and panel (c) is reprinted with permission from Reference \citenum{Gu2017} Copyright 2017 by the
American Physical Society.
}
\label{fig2}
\end{figure*}

\subsubsection{Polar Semiconductors}
An obvious way of obtaining polar metals is to dope polar insulators or semiconductors, such as wurtzite-structure (space group $P6_3mc$) GaN or ZnO. Both are direct band gap  
semiconductors and have been widely investigated in their doped states for their potential applications in short-wavelength
optoelectronic devices \cite{Ozgur2005}. $n$-type doping, with high conductivity can be readily achieved; in fact, undoped ZnO is a native $n$-type semiconductor due to intrinsic defects \cite{Ozgur2005,Ueno2019}. While the doped wurtzite structures therefore clearly fall into the polar metal category, to the best of our knowledge none of the associated exotic properties that we describe in Section~\ref{Section:PandP} have yet been measured; such investigations would be of great interest.

\subsubsection{Conventional Ferroelectrics}
It is less obvious that charge can be introduced into a conventional FE, for which transformation to the centrosymmetric parent phase is possible, without destroying the polar behavior. In particular, prototypical FEs such as perovskite BTO do not meet the Anderson-Blount criteria; electronic carriers populate the formally empty transition-metal Ti $d$ band, destroying the ``$d^0$''-ness that favors off-centering through the second-order Jahn-Teller effect, as well as screening the dipole-dipole interaction.
$n$-type doping has been achieved in  BTO through substitution of Ba$^{2+}$ by small amounts of La$^{3+}$ (or other rare earths), Ti$^{4+}$ by Nb$^{5+}$ and by introduction of oxygen vacancies, with the goal of making FE semiconductors for applications 
such as piezoelectric transducers, multilayer capacitors, and sensors with positive temperature coefficient of resistivity \cite{Jaffe1958,Moulson2003,Buscaglia2000,Chen2011,Kolodiazhnyi2003}. 
Semiconducting (possibly polaronic) behavior is established \cite{Iguchi1991,Gillot1992,Kolodiazhnyi2003,Kolodiazhnyi2008} with the FE distortion reported to persist in the conducting regime up to a limit of $\sim$0.1 electrons per unit cell  \cite{Wang2012,Michel2021,Kolodiazhnyi2010, Cordero2019} (see \textbf{Figures \ref{fig2}}\textit{a} and \textit{b}), or even higher values under compressive strain \cite{Chao2018}. Note that this limit is far smaller than the 0.75 electrons per W found computationally to be needed to suppress the W off-centering in antiferroelectric WO$_3$ \cite{Walkingshaw/Spaldin/Artacho:2004}, pointing to the importance of the screening of the dipole-dipole interaction in the FE case \cite{Wang2012}.

Interestingly, the polarization is much more robust to doping computationally with a uniform background {\it positive} charge (\textbf{Figure \ref{fig2}}\textit{b}) with an increase in polarization at small hole concentrations caused by weakening of the in-plane Ti-O bonds that need to bend to allow the Ti ions to displace away from their inversion centers. In addition, computational doping with impurity atoms can sometimes increase the ferroelectricity, since distortions caused by the size of the dopant ion and changes in lattice constants also play a role  \cite{Michel2021}. 
Note that the high aliovalent dopings studied computationally are likely unattainable experimentally, limited by the compensation of charge, for example, by cation vacancies, free electrons, or the change in the Ti valence state from Ti$^{4+}$ to Ti$^{3+}$ \cite{Lewis/Catlow:1996,Liu2020}; 
in particular there are few reports of robust $p$-type conductivity in BTO. 
Note also that neutron diffraction studies have suggested that phase separation between the polar and metallic phases might occur in some cases \cite{Jeong2011}.

\subsection{Adding Carriers To Unconventional Ferroelectrics}

Next, in the spirit of Anderson and Blount \cite{AndersonBlount1965}, we discuss FEs in which the atomic orbitals forming the top of the valence band or bottom of the conduction band are not key in driving the FE instability. In these cases, the ferroelectricity is likely to be more robust to metallicity. Indeed, a recent computational survey \cite{Zhao2018} identified a number of FE classes in which the polarization is insensitive to the introduction of carriers. We review these here and discuss additional directions for future exploration.

\subsubsection{Lone-Pair Active Ferroelectrics}
In lone-pair active FEs, the FE distortion is driven by the stereochemically active lone pairs of electrons which localize and displace their host cations introducing a polarization. Examples include the $6s^2$ lone pairs on the Pb$^{2+}$ ions in PTO or the Bi$^{3+}$ ions in BiFeO$_3$, and the $4s^2$ lone pairs on the Ge$^{2+}$ ions in GeTe.  The off-center distortions and their alignment are likely to be more robust to carriers than in $d^0$-driven off-centering, since the relevant lone-pair energy levels tend to be far from the Fermi energy. Computationally, this is established in the prototypical lone-pair active FE PTO, for which doping suppresses the $d^0$ Ti$^{4+}$ ion's off-centering but even enhances that of the Pb$^{2+}$ when lattice relaxations are included \cite{He2016,Zhao2018}. Experimentally, PbTi$_{1-x}$Nb$_x$O$_3$ has been shown using high-resolution
scanning transmission electron microscopy (STEM) to maintain its polar ground state up to $x = 0.12$ (\textbf{Figure \ref{fig2}}\textit{c}), at which point metallic conduction also occurs \cite{Gu2017,Takashi2000}, although the polarization versus electric field switching 
loops start to resemble those of lossy dielectrics by this doping level. Interestingly, domain boundaries with strong polar discontinuities are reported, likely reflecting the screening ability of the introduced carriers; it will be important to establish in future work whether the metallicity occurs in the bulk of the domains or is confined to these boundaries, as has been observed in the lone-pair active multiferroic, BiFeO$_3$ \cite{Seidel_et_al:2009}.

\subsubsection{Geometric Ferroelectrics} 

\begin{figure*}[t]
\includegraphics[width=\columnwidth]{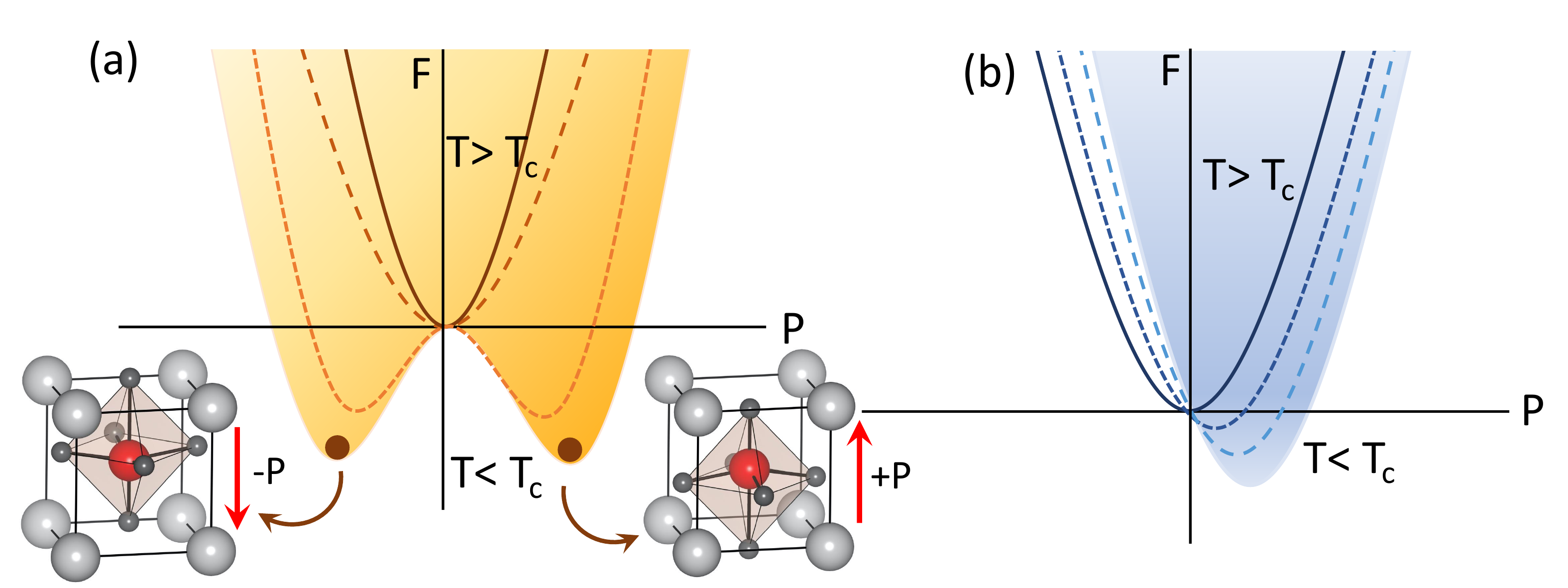}
\caption{Free energy across a PE to FE phase transition for (a) proper and (b) improper FEs. In proper FEs, the free energy in the FE phase has the characteristic double-well potential shown, and the polarization can be switched between its values $\pm P$ at the two minima of the well using an external electric field. In improper FEs, the polarization is shifted to a non-zero value through coupling to a non-polar primary order parameter, which is often a rotational (or combination of) distortion(s). }
\label{fig3}
\end{figure*}

In geometric FEs the symmetry lowering to the polar structure is caused by a structural collapse -- often consisting of rotations of the anionic polyhedra -- resulting from a mismatch between the ionic sizes, for example if the cations are too small to fill the spaces between the anionic polyhedra (see \textbf{Figure \ref{fig2}}\textit{d}). There is minimal chemical bond formation across the transition, and second-order Jahn-Teller considerations are not relevant, suggesting that metallicity might not be unfavorable. Geometric FEs can be either proper, in which the primary instability is itself polar (\textbf{Figure \ref{fig3}}\textit{a}), or improper, in which a secondary polar mode couples to one or more primary non-polar geometric instabilities (\textbf{Figure \ref{fig3}}\textit{b}).  Examples of such proper geometric FEs include La$_2$Ti$_2$O$_7$ \cite{LopezPerez/Iniguez:2011} and Ba$M$F$_4$ \cite{Ederer/Spaldin_2:2006}, where $M$ is a transition metal. Improper geometric FEs include the hexagonal manganites such as YMnO$_3$, as well as so-called hybrid improper (with two rotational instabilites) FEs such as the Ruddlesden-Popper (RP) phase Ca$_3$Mn$_2$O$_7$ \cite{Benedek/Fennie:2011}. In most geometric FEs the anionic polyhedra are layered, since three-dimensional connectivity does not generally yield polar rotation patterns. 

\paragraph{Doping Geometric Ferrroelectrics}
To our knowledge, the only studies of the effect of doping on geometric FEs to date have been computational. For  proper geometric FEs, the behavior of the  Carpy-Galy phases  La$_2$Ti$_2$O$_7$ \cite{LopezPerez/Iniguez:2011} and Sr$_2$Nb$_2$O$_7$ has been calculated \cite{Zhao2018} and in both cases the amplitude of the single polar rotational distortion responsible for their FE ground state \cite{NunesValdez/Spaldin:2019} was found to be insensitive to either electron or hole doping \cite{Zhao2018}. Analogous calculations for the hybrid improper RP FEs Ca$_3$Ti$_2$O$_7$ \cite{Zhao2018} and Sr$_3$Sn$_2$O$_7$ \cite{LiBirol2021} also found no suppression, and even an enhancement in the latter case. Calculations for the improper FE hexagonal manganites, and of course experimental attempts to verify the predictions, would be interesting directions for future work.
  
\paragraph{Geometric Polar Metals}

\begin{figure*}[t]
\includegraphics[width=\columnwidth]{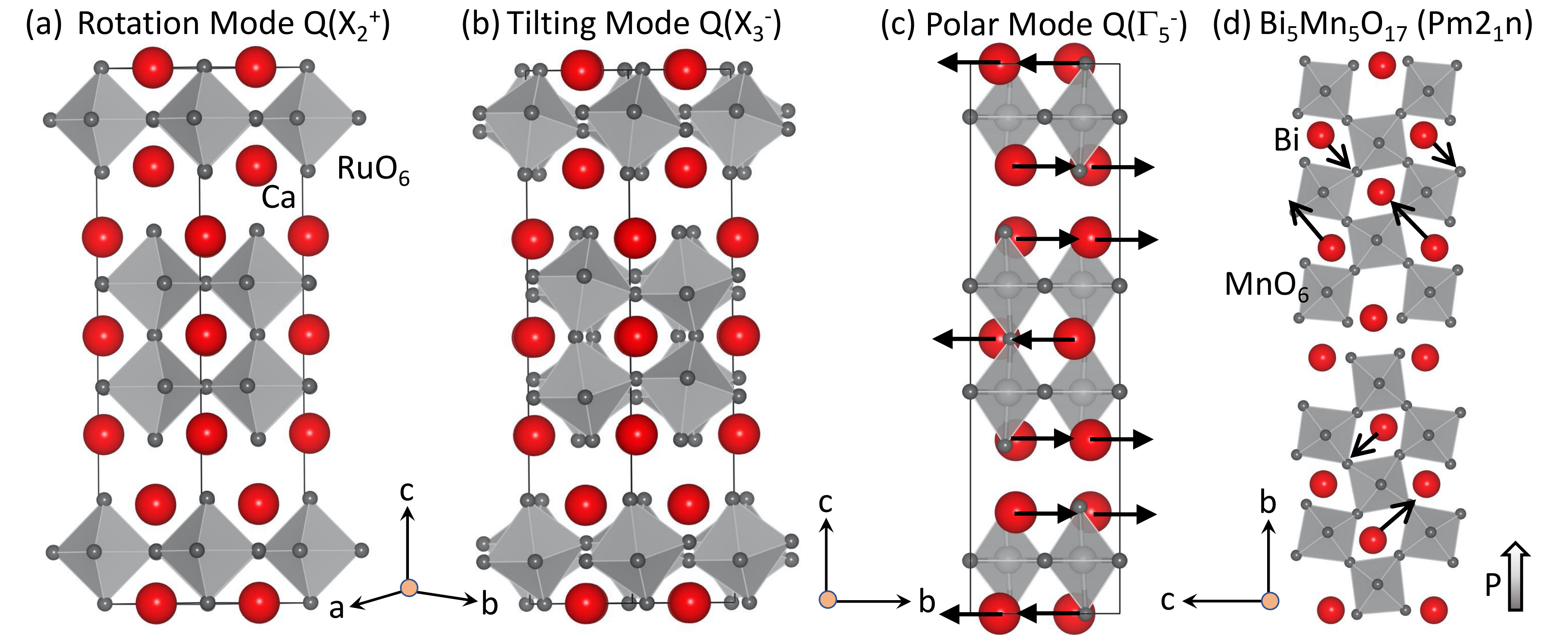}
\caption{Geometric polar metals. In Ca$_3$Ru$_2$O$_7$, the combined zone-boundary rotation (a) and tilt (b) modes couple to the zone center polar mode (c) to cause the polar symmetry. (d) The polar $Pm2_1n$ structure of Bi$_5$Mn$_5$O$_{17}$, depicting the displacements of the Bi atoms (thin arrows) that results in a net polarization P along $\vec b$ direction. Panels (a)-(c) are adapted with permission from Reference \citenum{Lei2018} Copyright  2018 American Chemical Society. Panel (d) is adapted from Reference \citenum{Urru2020}.
}
\label{fig4}
\end{figure*}

While metallic doping of geometric FEs has not to our knowledge been achieved in experiment, polar metals, in which the polar symmetry occurs because of the same geometric structural distortions, are established. The prototype is the $n=2$, RP structure Ca$_3$Ru$_2$O$_7$, which has the same polar structure as Ca$_3$Mn$_2$O$_7$ \cite{Yoshida2005}: the zone-boundary rotation ($X_2^+$) and tilt ($X_3^-$) modes that couple to the zone-center ($\Gamma_5^-$) polar mode to give rise to the polar ground state (\textbf{Figure \ref{fig4}}\textit{a-c}) are unaffected by the Ru-$t_{2g}^4$ states that dominate the metallic conduction. Interestingly, 90$^{\circ}$ and 180$^{\circ}$ polar domains were recently imaged using optical second harmonic generation, and the 90$^{\circ}$ were switched using strain \cite{Lei2018}.  A related prediction is metallic (Sr,Ca)Ru$_2$O$_6$ \cite{PuggioniRondinelli2014}, in which coupling of the non-zone center $M_2^+$ and $M_5^-$ RuO$_6$ rotation modes and the zone-center polar $\Gamma_5^-$ distortion is proposed to result in a polar displacement of the Sr and Ca ions.  
 
 The metallic layered-perovskite niobate and titanate Carpy-Galy phases with general formula  A$_n$B$_n$O$_{3n+2}$, such as SrNbO$_{3.45}$ ($n=4.5$) and SrNbO$_{3.41}$ ($n=5$), are also polar metals, in which quasi-1D metallic conductivity coexists with polarity \cite{Lichtenberg20011}. Future explorations of FE switching in these materials are to be encouraged. 
Density functional computations indicate that incorporation of lone-pair active cations, such as in $n=5$ Bi$_5$Ti$_5$O$_{17}$ would enhance the polar distortions \cite{Filippetti2016},
and suggest that a multiferroic metal could be engineered by replacing Ti with Mn (\textbf{Figure \ref{fig4}}\textit{d}) \cite{Urru2020}.

\subsection{Artificially Layered And Two-Dimensional Systems}

We saw in the previous section that the naturally layered geometric FEs and their metallic analogues provide a promising platform for designing polar metals. Next, we discuss artificially layered structures, engineered for example by thin-film growth techniques, as well as naturally two-dimensional systems. These offer the possibility of planar metallicity combined with perpendicular polarization in a capacitor-like geometry, and so are particularly good candidates for achieving switching of the polarization. 

The most prominent example of a 2D FE metal is WTe$_2$, which we discuss in detail in the next section; related transition-metal dichalcogenides are ripe for exploration. Other examples include proposals of CrN and CrB$_2$ monolayers \cite{Luo2017}, which are predicted to be {\it hyperferroelectric} (see section~\ref{hyperferroelectric}).  

In thin-film heterostructures, various clever engineering approaches have been employed. For example, polar distortions have been induced in ultra-thin (three-unit-cell thick) metallic SrRuO$_3$ by sandwiching it between layers of FE BTO \cite{Meng2019} without destroying its metallicity. Interestingly, here the resulting polar metal is also magnetic.  In the usually centrosymmetric metallic rare-earth nickelates, such as NdNiO$_3$ and LaNiO$_3$, a polar displacement of the rare-earth ions has been stabilized by modifying the NiO$_6$ octahedral rotation pattern through proximity to a LaAlO$_3$ substrate.
Finally, artificial layering has been used successfully to introduce carriers into conventional FEs via field effects from a polar discontinuity. Examples include Ba$_{0.2}$Sr$_{0.8}$TiO$_3$/LaAlO$_3$ heterostructures, in which a 2D electron gas forms in the Ba$_{0.2}$Sr$_{0.8}$TiO$_3$ interfacial region while the polarity is preserved \cite{Zhou2019}; a similar mechanism has been demonstrated in BaTiO$_3$/SrTiO$_3$/LaTiO$_3$ superlattices \cite{Cao2018}. 

\subsection{Leading Candidates For FE-Like And FE Metals}
\subsubsection{ LiOsO$_3$: FE-Like Metal}

\begin{figure*}[t]
\includegraphics[scale=0.4]{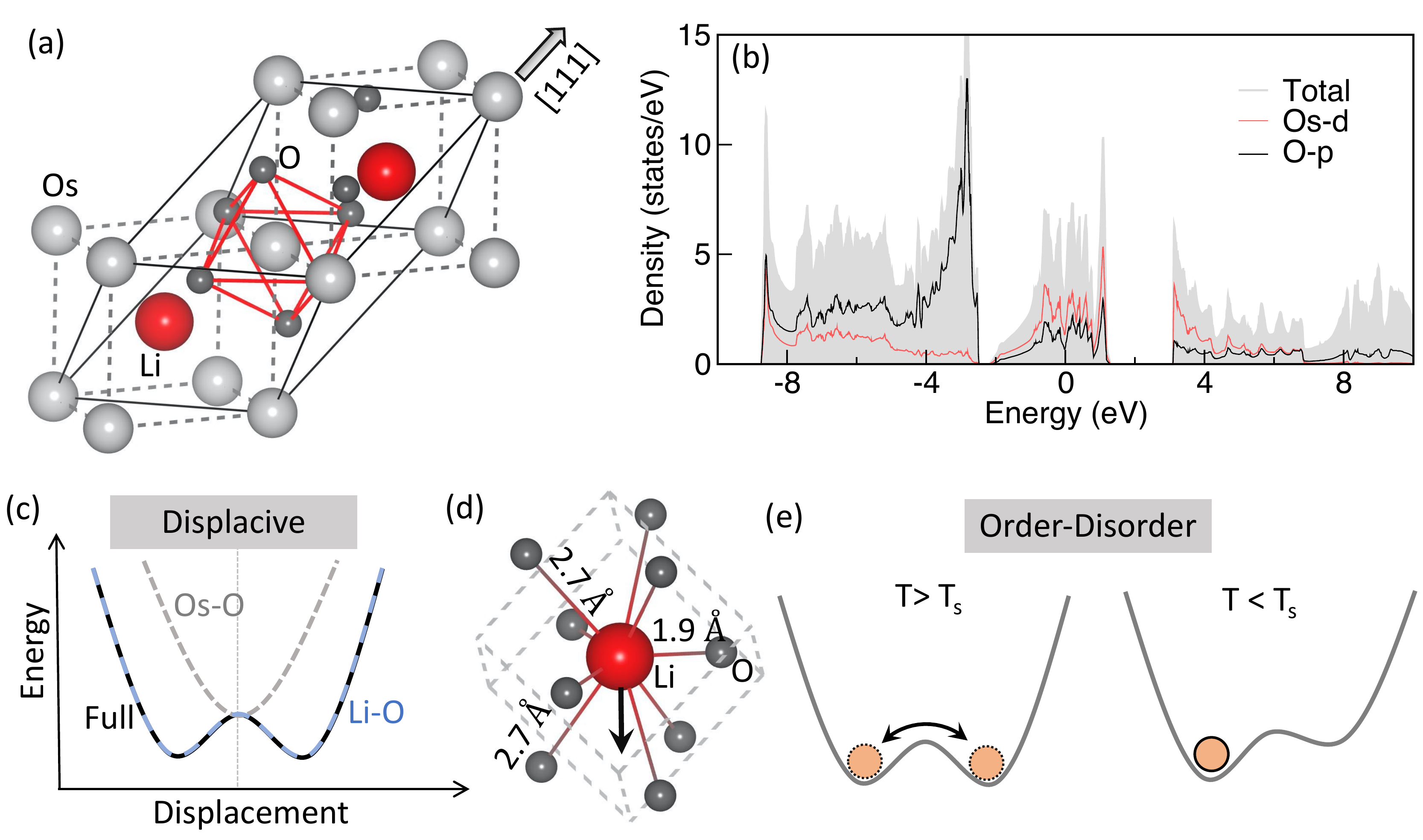}
\caption{FE-like polar metal LiOsO$_3$. (a)  The high-temperature $R\bar 3c$ crystal structure of LiOsO$_3$, related to the cubic $Pm$-$3m$ structure by out-of-phase ($a^-a^-a^-$) octahedral rotations around the cubic [111] axis. The $R\bar 3c$ unit cell is shown with the solid black line, and the pseudo-cubic unit cells with dashed lines. In the polar $R3c$ structure the Li ions displace along the cubic [111] direction indicated.
(b) The computed total and partial densities of states, showing that the Fermi energy (set at zero) is dominated by Os-$d$ states strongly hybridized with O-$p$ states. (c) Schematics of the potential energy vs displacement, illustrating that the unstable polar phonon mode at the $\Gamma$ point is driven by the displacement of the Li atoms with respect to the O atoms, suggesting a displacive structural transition. (d) The geometric origin of the Li-atom displacement. The large out-of-plane Li-O distances compared to the in-plane distances lead to tolerance factor $t_R < 1$ and cause the displacement of the Li atom in the out-of-plane direction indicated in black arrow. The dashed line indicates the cubic unit cell shown in (a). (e) Schematics of the Li potential well diagram, depicting the proposed order-disorder nature of the structural transition in LiOsO$_3$.  Panel (c) is reprinted with permission from Reference \citenum{Xiang2014} Copyright 2014 by the
American Physical Society.}  
\label{fig5}
\end{figure*}

The origin of electric polarization in the established FE insulating perovskites Li$M$O$_3$ with $M=$Nb or Ta is somewhat unconventional in that, in addition to the formally $d^0$, $M^{5+}$ ions, a ‘rattling’ of the small, underbonded Li$^+$ ions contributes to the polarization. Molecular-dynamics simulations \cite{Phillpot2004} even suggest a two-stage FE phase transition in LiNbO$_3$, with an order-disorder transition in the Li–O planes followed by a displacive transition in the Nb–O cages at lower temperature. Isostructural LiOsO$_3$ was first achieved in 2013 by high-pressure synthesis \cite{Shi2013}. It has a continuous second-order phase transition from non-polar $R\bar 3c$ to polar $R3c$ symmetry at $T_s=$ 140 K (see \textbf{Figure \ref{fig5}}\textit{a}) which, in the absence of  $d^0$-ness on the Os$^{5+}$ ion, is solely driven by the ‘rattling’ of the Li ions. Combined with its conducting behavior from the partially filled Os-$t_{2g}$ states strongly hybridized with the O-$p$ states at  the Fermi energy (see \textbf{Figure \ref{fig5}}\textit{b}), it provides the first unambiguous example of a ‘FE-like’ metal, although its residual resistivity is around two orders of magnitude larger than that of a prototypical metal. Its behavior is consistent with the decoupled mechanism of Anderson and Blount \cite{AndersonBlount1965,PuggioniRondinelli2014}, since the conductivity arising from  the Os-$t_{2g}$ bands is largely decoupled from the Li-ion driven polar instability. The discovery of LiOsO$_3$ has led to a a flurry of research work aiming to understand the microscopic origin of the FE-like phase transition, in particular whether it is displacive or order-disorder, as well as the mechanism for long-range ordering of the local electric dipoles.  
\paragraph{Origin Of Polar Instability: Displacive Versus Order-Disorder} 
The FE-like phase transition from $R\bar 3c$ to $R3c$ at $T_s$ in LiOsO$_3$ can be described by condensation of an unstable zone-center polar $A_{2u}$ phonon mode, which has been detected in ultra-fast optical pump-probe measurements \cite{Laurita2019}. 
DFT indicates that the resulting structural distortion consists of Li displacements of $\sim$ 0.5 \AA~along the pseudocubic [111] direction (\textbf{Figure \ref{fig5}}\textit{a}) accompanied by Os-O bending \cite{Xiang2014}. The double-well energy profile of the full $A_{2u}$ phonon mode is almost completely captured by the Li-O displacements, with the Os-O displacements alone actually increasing the energy from the non-polar structure (\textbf{Figure \ref{fig5}}\textit{c}) suggesting that the FE-like instability is driven primarily by the Li atoms \cite{BenedekBirol2016}, although O-$p$ hybridization with empty Os-$e_g$ states likely also plays a role  \cite{GiovannettiCapone2014}. 
This is consistent with the small size of the Li ion relative to its oxygen coordination cage (\textbf{Figure \ref{fig5}}\textit{d}), causing it to shift out of the plane of its three closely coordinated oxygen neighbors towards three of the six distant oxygen atoms (see
\textbf{Figure \ref{fig5}}\textit{d}) to relieve its underbonding \cite{Xiang2014}. While the soft polar mode points to a displacive transition, anomalous increases in Raman mode linewidths across $T_s$ on heating \cite{Jin2016,Feng2019} as well as incoherent charge transport  above $T_s$ \cite{Shi2013} suggest order-disorder character. Indeed, the low transition temperature ($T_s$ = 12 meV) compared with the computed depth of the double-well potential (44 meV) \cite{Liu2015} points to the order-disorder picture as sketched in \textbf{Figure \ref{fig5}}\textit{e}. As in conventional FEs, it is likely that the polar phase transition in LiOsO$_3$ has aspects of both displacive and order-disorder nature, which are variously emphasized depending on the time or length
scale of the probe \cite{Laurita2019,Sim2014}. 

A consequent question is what drives the Li ions to all off-center in the same direction instead of, for example, in an anti-aligned  pattern that would lead to an overall non-polar structure. 
Model Hamiltonian and Monte Carlo simulations suggest that long-range electrostatic forces do not play a role in the parallel alignment of the Li ions \cite{Xiang2014}, with DFT calculations indicate that the mode with alternating displacements of the Li atoms is simply less unstable than the polar mode  \cite{BenedekBirol2016}. Indeed, condensation of this anti-polar mode would lead to a non-polar $R\bar 3$ structure with face-shared Os-centered and Li-centered octahedra, which are associated with reduced stability according to Pauling’s third rule \cite{Pauling1960}. 
The highly anisotropic incomplete screening of local dipoles resulting from the strongly correlated bad metallic behavior with its small quasi particle weight \cite{Vecchio2016} may also play a role  \cite{Liu2015}. 

Clearly the discovery of FE-like LiOsO$_3$ has been a key step in the development of the field of polar metals. Important ongoing work includes tuning of the phase transition temperature, with a $T_s$ of 250 K achieved with a pressure of 6.5 GPa \cite{Paredes2018}, and DFT predictions of enhanced $T_s$ using compressive strain. Conversely, tensile strain is predicted to favor the non-polar state, suggesting a strain-driven quantum phase transition at around 2.5$\%$ tensile strain \cite{Narayan2019} which awaits experimental realization. While a switchable electric polarization with an asymmetric hysteresis loop has been predicted for two-unit cell thick LiOsO$_3$ \cite{Lu2019}, bulk switchability is likely to remain challenging due to the three-dimensional metallic character. In this context,  
the first discovery of a switchable FE metal in layered WTe$_2$ is an important development, which we discuss next. 
 
\subsubsection{WTe$_2$: FE Metal}

 \begin{figure*}[t]
\includegraphics[width=\columnwidth]{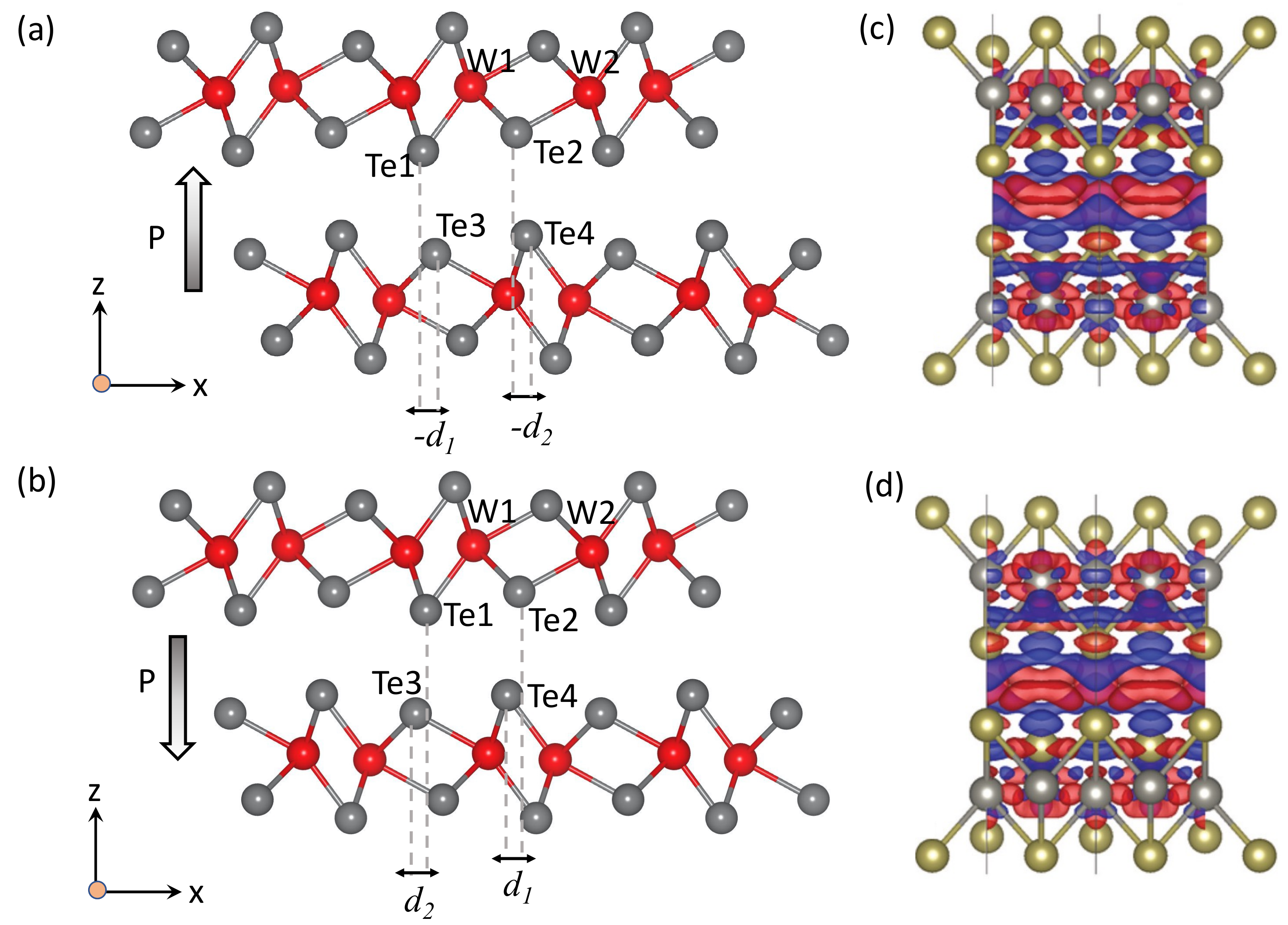}
\caption{FE metal WTe$_2$. 
(a)-(b) Switching of the electric polarization $P$ (indicated with large arrows) in WTe$_2$ through sliding of the upper layer by ($d_1+d_2$) along $-x$ direction. (c)-(d) Computed differential charge density for the structures in (a) and (b). 
Panels (a) and (b) are adapted with permission from
Reference \citenum{Yang2018} Copyright 2018 American Chemical Society, and panels (c) and (d) are used with permission of Royal Society of Chemistry, from Reference \citenum{Liu2019};
permission conveyed through Copyright Clearance Center, Inc.
}
\label{fig6}
\end{figure*}

The recent experimental demonstration of room-temperature switchable spontaneous out-of-plane polarization in bi- and tri-layered \cite{Fei2018}, and bulk WTe$_2$ \cite{Sharma2019} is a revolutionary step, establishing the first example of a FE metal.  
The $T_d$ phase of bulk WTe$_2$ is a polar orthorhombic layered structure with $Pmn2_1$ symmetry. Each layer consists of edge-shared WTe$_6$ octahedra and is centrosymmetric, with successive layers rotated by 180$^\circ$ relative to each other. It is metallic, with the metallicity, which is strongly confined to the planes  \cite{Sharma2019}, persisting down to a thickness of three layers. The absence of band dispersion along $\Gamma$ to $Z$ means that a spontaneous electrical polarization along the [001] direction is possible. While this has been calculated using DFT to be very small, $\sim$0.19 $\mu$C/cm$^2$ \cite{Sharma2019}, it has been switched and its domains have been imaged using piezoresponse force microscopy \cite{Sharma2019,Fei2018}.        

The polarization switching mechanism is unconventional. First, the computed energy barrier between two bulk structures of opposite polarizations related by inversion symmetry 
is prohibitively high. The hexagonal symmetry, however, allows for three equivalent distortion vectors, 
and the displacements required to switch between opposite polarization states of different distortion vectors are rather small with correspondingly lower energy barrier \cite{Sharma2019}. 
Second, the proposed atomic displacements involve a relative horizontal shifting of the layers, illustrated for the bilayer case in \textbf{Figure \ref{fig6}}\textit{a-b} \cite{Yang2018,Liu2019}. Shifting the relative positions of the layers along the $x$ direction so that the horizontal distances between the nearest interlayer Te ions change from $-d_1$ and $-d_2$ ($P\uparrow$, panel \textit{a}) to $+d_2$ and $+d_1$ ($P\downarrow$, panel \textit{b}) reverses the interlayer chemical bonding (\textbf{Figure \ref{fig6}}\textit{c-d}) and correspondingly the electric polarization in the vertical direction. 

This proposed inter-layer sliding mechanism has been referred to as a ``decoupled space mechanism" by analogy to the Anderson-Blount decoupled electron mechanism, since the vertical displacements of the localized electrons and the horizontal  inter-layer displacements and metallicity are spatially decoupled. While this model provides useful insight into the FE metallic behavior of WTe$_2$, some open questions remain, in particular whether the two-layer illustration of \textbf{Figure \ref{fig6}} can be extended to bulk WTe$_2$, whether such a mechanism for ferroelectricity should be classified as being of proper or improper type, and how the out-of-plane electric field couples to the in-plane inter-layer sliding.

\subsection{Promising New Design Directions}
Before closing this section, here we point out a few promising directions in designing polar metals for future investigations.
\subsubsection{Hyperferroelectrics}
\label{hyperferroelectric}
Hyperferroelectrics \cite{Garrity2014}, are an unusual class of proper FEs that have a persistent polarization even in the presence of an unscreened depolarizing field  \cite{Garrity2014,Bennett2012,Birol2018}. As a result they are good candidates for becoming polar metals upon doping. They are distinguished from conventional FEs by their small Born effective charges (the amount of polarization generated by an ionic displacement) and correspondingly weak LO-TO splittings, so that even their LO modes can be unstable. 
In a recent theoretical work \cite{Zhao2018} many layered geometric FEs, including the Carpy-Galy phases La$_2$Ti$_2$O$_7$, Sr$_2$Nb$_2$O$_7$, and Ca$_3$Ti$_2$O$_7$, were identified as hyperferroelectrics with weak LO-TO splitting and were predicted to be polar metals on doping. Several of the metallic counterparts of these hyperferroelectric structures have also been recently proposed  \cite{Sante2016,Chen2017,Luo2017,Gao2018,Du2019}.

 \subsubsection{Spin-Spiral-Driven Polar Metals} \label{spinspiral}
 
 An as-yet unexplored possibility for designing polar metals is to introduce magnetic spin spirals that break the inversion symmetry into metallic systems, analogous to the established spin-driven FE polarization in the prototypical type-II multiferroic, TbMnO$_3$ \cite{Kimura2003}. Although rare, several magnetic polar metals have already been reported \cite{Lei2019,McCall2003,Yuan2019,Zhang2022}. Among these, interestingly, Fe-doped Ca$_3$Ru$_2$O$_7$, Ca$_3$(Ru$_{0.95}$Fe$_{0.05}$)$_2$O$_7$, has an incommensurate spiral spin structure which resembles that of TbMnO$_3$ \cite{Lei2019}. While there might be no connection between the spin spiral and the polarity in this case (the parent compound  Ca$_3$Ru$_2$O$_7$ is already polar even in the absence of magnetism, and hosts an antiferromagnetic state below $T_N$= 56 K while remaining metallic), we see identification of metals in which the structural polarity originates solely from the spin structure as an exciting future direction.

\subsubsection{Modifying Conventional Metals To Make Them Polar} \label{ElementalPolarMetal}

As discussed in Section \ref{CentroInsulator}, conventional metals tend to be centrosymmetric to minimize the Coulomb repulsion between the ions. To make a regular metal polar, therefore, some degree of covalency is likely necessary to overcome the additional Coulomb repulsion. Indeed, this is the case in the noncentrosymmetric trigonal Te which is, however, {\it chiral} rather than polar.  In this context, an intriguing strategy has been recently proposed for designing {\it elemental} polar metals \cite{Zhang2018}. The approach requires at least two different Wyckoff positions (necessary because equivalent Wyckoff positions would result in a nonpolar space group), with at least one of them lacking inversion symmetry, occupied by atoms of the same element. 
Based on this idea, elemental metastable polar metal phases of group-V phosphorus, arsenic, antimony, and bismuth, with lone-pair driven structural transitions from non-polar $P6_3/mmc$ to polar $P6_3mc$ space groups were proposed \cite{Zhang2018} and await experimental verification.    

\section{PROPERTIES AND PROSPECTS}
\label{Section:PandP}
The experimental realization of polar metals 
\cite{Shi2013,Yoshida2005,Lei2018,Nukala2017,ShirodkarWaghmare2014,Fei2018,Sharma2019,Fei2016,PuggioniRondinelli2014,Ding2017}
has opened the door to exploration of many exotic quantum phenomena that had previously been inaccessible. While many interesting potentially useful properties of polar metals, such as, unconventional superconductivity \cite{Rischau2017}, thermoelectricity \cite{Lee2012,Sakai2016}, complex magnetism \cite{Lei2019,Zhang2022}, multiferroicity \cite{Luo2017}, and nontrivial topology \cite{Wang2016,Wu2016}, have been realized, many proposed behaviors are yet to be uncovered. Here, we summarize some properties and potential applications that we find particularly promising for future investigation.      

\subsection{Rashba Interaction and Spintronic Applications}

\begin{figure*}[t]
\includegraphics[width=\columnwidth]{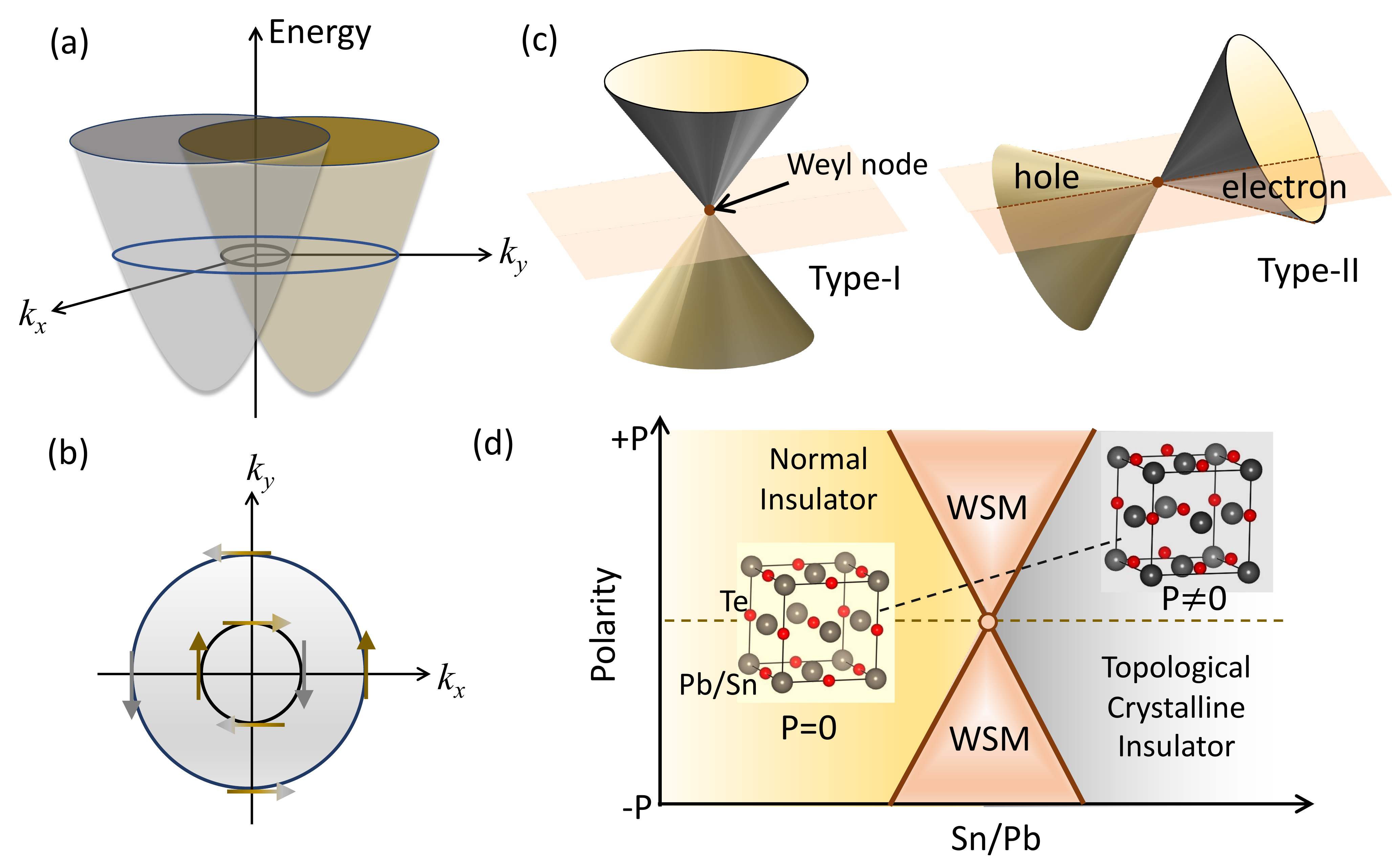}
\caption{(a) Schematics showing the splitting of the two spin-polarized bands (parabolic bands with different colors) in the Rashba effect. (b) Top view of the spin textures at the energy contours shown in (a). (c) Illustration of type-I (\textit{left}) and type-II (\textit{right}) WSMs. (d) Schematic phase diagram, adapted from Reference \citenum{Zhang2021}, showing topologically distinct phases of Pb$_{1-x}$Sn$_x$Te alloy with the variation of two parameters, polarity and Sn/Pb ratio.   
}
\label{fig7}
\end{figure*}

Polar/FE metals provide a natural platform for investigating the Rashba-type interaction, the key ingredients of which are the broken inversion symmetry that is intrinsic to polar/FE metals, and spin-orbit interaction, which can be enhanced by choosing appropriate polar metals with heavy elements. The typical form of the Rashba interaction is
\begin{equation} \label{Rashba}
 {\cal H}_R=\alpha_R(\vec \sigma \times \vec k)\cdot \hat z, 
\end{equation}
where the Rashba coefficient $\alpha_R$ depends on the strength of the spin-orbit coupling, and $\hat z$ denotes the direction of the polar axis (for the extrinsically broken inversion symmetry case, it corresponds to the direction of the applied field that breaks the symmetry). Such an interaction ${\cal H}_R$ results in a linear-$k$ splitting between the two oppositely spin-polarized bands $E_{\pm} (k)$ on top of the usual (e.g.\ parabolic) band dispersion, $E_{\pm} (k) = \frac{\hbar^2k^2}{2m^*} \pm \alpha_R k$ as depicted in \textbf{Figure \ref{fig7}}\textit{a}. 

Interest in the Rashba effect primarily stems from the resulting spin texture in momentum space (see \textbf{Figure \ref{fig7}}\textit{b}) that is clear from rewriting the Rashba Hamiltonian in Eq. \ref{Rashba} as ${\cal H}_R=\vec B_k\cdot \vec \sigma$. Here $\vec B_k = \alpha_R(\vec k \times \hat z)$ acts as a momentum-dependent magnetic field that defines the electron spin direction $\vec \sigma$ in the momentum space $\vec k$. This spin-momentum locking of the Rashba interaction is one of the key ingredients for spintronic applications, where a spin polarization is generated from a charge current and vice-versa (for an excellent review on the spintronic applications of the Rashba effect, see Reference \citenum{Manchon2015}).

Interestingly, switching the polarization direction ($\hat z$) of a FE metal, also switches the direction of the magnetic field  $\vec B_k$ and hence the spin orientation in the momentum space, providing an additional tunability to the spin texture. 
Conventional FEs with heavy elements are proposed to be good candidate materials \cite{Sante2013,Arras2019}, broadening the horizon of the Rashba effect from surfaces and interfaces \cite{Ishizaka2011,Shanavas2014,Bhowal2019} 
 to bulk FE Rashba semiconductors with switchable spin texture \cite{Picozzi2014}.
In this context, recent measurements \cite{Varotto2021} showing FE switching in GeTe thin films by electrical gating even in the presence of high carrier density is particularly promising. This work demonstrates that Fe/GeTe heterostructures are efficient spin-charge converters, with a normalized charge current magnitude (measured in spin-pumping experiments) comparable to that of Pt, and the additional advantage of reversing the spin-charge conversion coefficient by changing the polarization direction. FE metals further enhance these possibilities due to the existence of an electric current without needing to introduce doping. 

\subsection{Topological Semimetals}
Analogues to the well known Dirac and Weyl fermions in the standard model of relativistic high-energy physics are well established in the low-energy excitations of condensed matter systems such as topological semimetals (TSMs) \cite{YanFelser2017,Armitage2018,Lv2021}.
Recently, many polar metals have been predicted to host such topological Dirac and Weyl points in their electronic band structure offering the additional possibility of controlling/manipulating the topological properties using electric fields. Here we discuss the connection between polar metals and TSMs. 

\subsubsection{Weyl Semimetals}
Weyl semimetals (WSMs) are characterized by protected 
two-fold degenerate band crossings in momentum space near the Fermi level, leading to zero-dimensional Weyl nodes (see e.g., \textbf{Figure \ref{fig7}}\textit{c}) that appear in materials which lack either space-inversion or time-reversal symmetry. In the WSMs that lack time-reversal symmetry, the WSM phase is magnetic and, therefore, the corresponding intriguing electronic and optical properties can be tuned by manipulating the magnetic order.
Conversely, space-inversion-symmetry-broken WSMs, for which polar metals are good candidates, should allow for control of the Weyl nodes near the Fermi level using an electric field. 

To the best of our knowledge such non-magnetic WSM behavior was in fact first observed in the polar metal TaAs \cite{Xu2015,Lv2015,Huang2015}. Even more exciting are {\it ferroelectric} non-magnetic WSMs, such as the recently discovered In-doped Pb$_{1-x}$Sn$_x$Te alloy \cite{Zhang2021}. While in the rock-salt phase, PbTe is a normal insulator and SnTe is a topological crystalline insulator \cite{Tanaka2012}, at critical concentration $x_c \sim 0.35$, the mixed phase Pb$_{1-x}$Sn$_x$Te undergoes a transition from a normal to a topological crystalline insulator \cite{Dziawa2012}. At low temperature, SnTe undergoes a FE transition with polarization along the $<111>$ direction \cite{ONeill2017}, whereas PbTe is an incipient FE. Using in-plane Nernst effect measurements, Zhang {\it et. al.} \cite{Zhang2021} showed that for a specific Pb/Sn composition range and In doping concentration, it is possible to retain the ferroelectricity (with the polarization along $<001>$) and the non-trivial topology in the presence of finite carrier concentration, as illustrated in \textbf{Figure \ref{fig7}}\textit{d}.
A similar FE Weyl semimetal phase has been predicted for HgPbO$_3$ \cite{Li2016} and awaits verification. Likewise, the calculated electronic structures of hexagonal ABC-type polar metals (the metallic counterpart of the {\it hyperferroelectrics}) host intriguing Dirac and Weyl fermions and are promising materials for future experimental study \cite{Sante2016,Chen2017,Gao2018}.

\subsubsection{Type-II Weyl Semimetals}
Interestingly, the first FE metal, WTe$_2$, was well-known as a so-called type-II WSM, before its ferroelectricity was discovered. In contrast to the conventional WSMs (also known as type-I WSMs) discussed above, which have point-like Fermi surfaces, type-II WSMs have a characteristic tilted linear-in-momentum cone-like energy spectrum (\textbf{Figure \ref{fig7}}\textit{d}) with the Weyl points appearing at the contact of electron and hole pockets. The eight such type-II Weyl points in WTe$_2$, half located at 0.052 eV and half at 0.058 eV above the Fermi energy, were first predicted computationally \cite{Soluyanov2015} and later confirmed using angle-resolved photoemission spectroscopy \cite{Wang2016,Wu2016} and magneto-transport \cite{Peng2017} measurements.  Note that type-II Weyl fermions in condensed matter systems do not have a high-energy analogue as their characteristic tilt breaks Lorentz symmetry. 

\subsubsection{Dirac Points And Nodal Rings}
Interesting topological features have also been reported in the band structure of LiOsO$_3$, with striking changes  between the centrosymmetric and noncentrosymmetric structures \cite{Yu2018}. Yu {\it et. al} showed that the band structure of non-polar LiOsO$_3$ hosts multiple linear Dirac points (four-fold degenerate band crossings with linear dispersion) and a cubic Dirac point (four-fold degenerate band crossing with cubic dispersion in a plane but with linear dispersion in the other direction) at the T point in the Brillouin zone \cite{Yu2018}, stabilized by the nonsymmorphic glide mirror symmetry of the crystal structure.  
The absence of inversion symmetry in the polar $R3c$ structure removes the degeneracy of the Dirac points, transforming each linear Dirac point at L into a nodal ring on the glide mirror plane enclosing the L point (\textbf{Figure \ref{fig8}}\textit{a} and \textit{b}) and the cubic Dirac point into three nodal rings that cross each other. Such linear and cubic Dirac points and nodal rings may give rise to negative magnetoresistance, unusual quantum interference contributions to magnetoconductivity or highly anisotropic electrical transport \cite{Lv2021}.

To summarize this section, reconciling ferroelectricity and metallicity with non-trivial band topology offers the rich plethora of  Weyl fermion physics, combined with the additional electric-field control of FE materials. This suggests that exotic Weyl phenomena such as the chiral magnetic effect, or oscillations in thermal conductivity driven by chiral zero sound \cite{SongDai2019,Xiang2019}, could be activated by switching on and off the Weyl phase or tuned by controlling the separation of the Weyl points in momentum space. 
 Conversely, WSMs provide a promising platform for hosting new polar metal phases \cite{Li2016}, because the charge carriers in Weyl nodes are confined on the orbitals that form the linearly dispersive bands. Consistent with the original Anderson and Blount philosophy \cite{AndersonBlount1965}, such confined charge carriers are advantageous over the parabolic band dispersions in normal metals that allow the electrons to travel over the whole lattice. Furthermore, the much smaller charge carrier density of WSMs compared to  normal metals indicates weaker electrostatic screening, again favoring the polar metal phase.

\subsection{Non-Linear Hall Effect and Berry Curvature Dipole} \label{NHE}
The non-linear Hall effect (NHE) describes the generation of a second-order Hall current $j^{2\omega}$ in response to an applied electric field $\vec {\cal E}$, $j_a^{2\omega}=\chi_{abc} {\cal E}_b {\cal E}_c$, where $\chi_{abc}$ is the non-linear conductivity. It occurs in noncentrosymmetric metals that are also gyrotropic \cite{SodemannFu2015}, and since all polar point groups are gyrotropic, it is allowed by symmetry in all polar metals. Moreover, in FE-like metals, the onset of the NHE should mark the centrosymmetric to polar structural transition.

\begin{figure*}[t]
\includegraphics[scale=0.4]{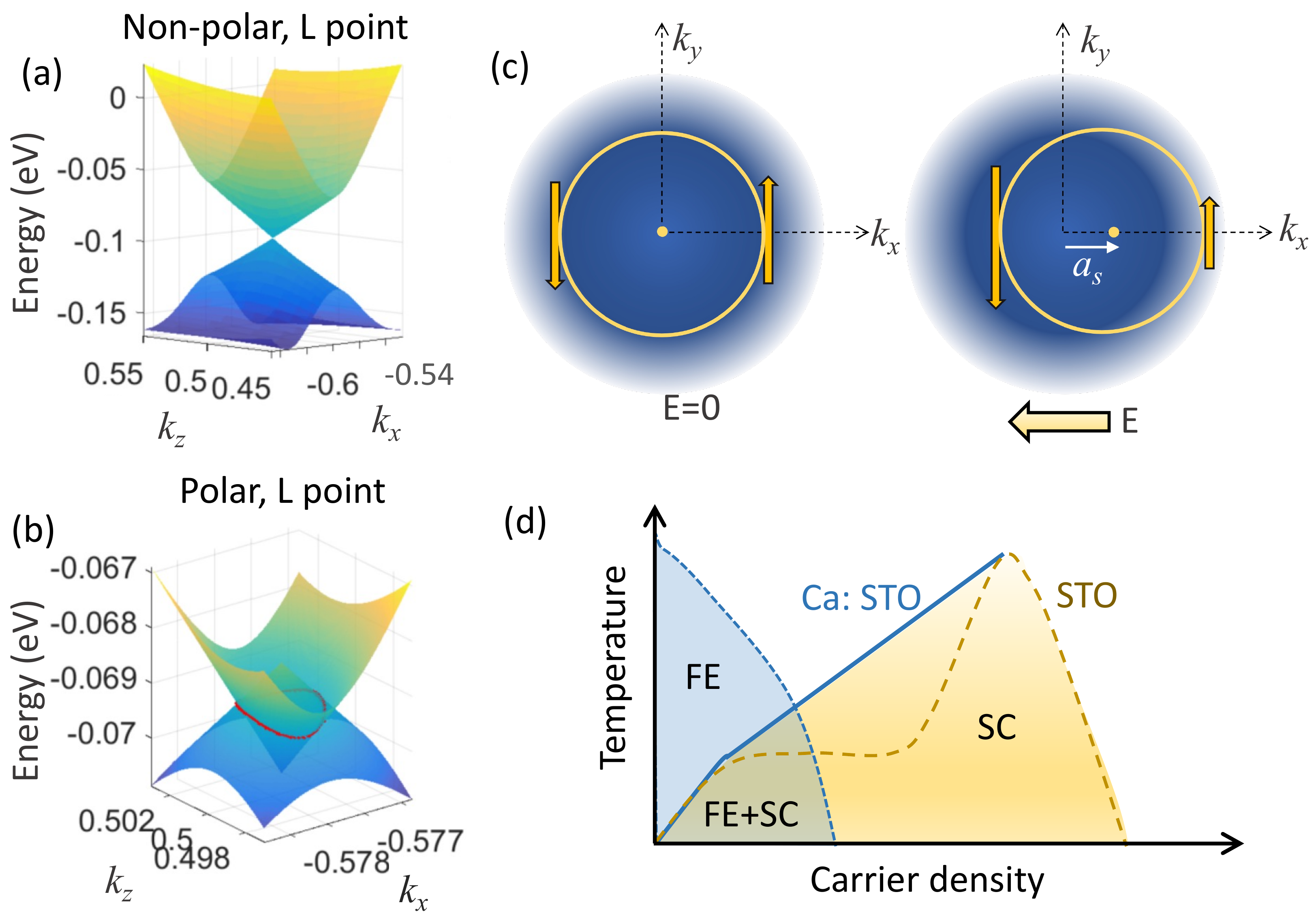}
\caption{Band dispersion around the L point in the Brillouin zone for (a) non-polar and (b) polar phases of LiOsO$_3$. 
(c) Schematics showing absence of net Berry curvature in absence of an electric field $\vec E$ (\textit{left}) and incomplete cancellation of the Berry curvature when a field $\vec E$ is applied (\textit{right}).
The shaded region shows the distribution of the Berry curvature (magnetic moment) with the dark and light colors denoting large and small magnitudes. The circles denote the Fermi surface, and the arrows show the direction of the Berry curvature (magnetic moment) at representative $k$ points.  The shift $\vec {a}_s$ of the Fermi surface in presence of $\vec E$ leads to incomplete cancellation of the Berry curvature at those representative points, resulting in a net Berry curvature (magnetization) at non-equilibrium. (d) Schematic phase diagram,  showing the coexistence of FE and superconducting (SC) order and the increase in SC T$_c$ (solid line) upon Ca substitution.
Panels (a) and (b) are reprinted with permission from Reference \citenum{Yu2018} Copyright 2018 by the
American Physical Society, and panel (d) is adapted by permission from Springer Nature Customer Service Centre GmbH: Springer Nature, Reference \citenum{Rischau2017}, \copyright 2017.
}
\label{fig8}
\end{figure*}

The origin of the NHE in polar metals is their underlying Berry curvature dipole. In a nonmagnetic system at equilibrium, the net Berry curvature $\vec \Omega(\vec k)$ is zero due to time-reversal symmetry that dictates $\vec \Omega(\vec k) \overset{\cal T}{\rightarrow} -\vec \Omega(-\vec k)$. However, such cancellation of $\vec \Omega$ at $\pm \vec k$ is incomplete in an electric-current-induced non-equilibrium electron distribution provided that the crystal structure is gyrotropic (\textbf{Figure \ref{fig8}}\textit{c}). The resulting net $\vec \Omega$ as a second order response to $\vec {\cal E}$ gives rise to the NHE with the conductivity 
 $\chi_{abc}=-\varepsilon_{adc} \frac{e^3\tau}{2(1+i\omega\tau)} {\cal D}_{bd} $, where $\varepsilon_{adc}$, $\tau$, $e$ are respectively the Levi-civita symbol, relaxation time-constant, and the electronic charge, and ${\cal D}_{bd}$ is the Berry curvature dipole, defined as,
\begin{equation}
  {\cal D}_{bd}=\frac{1}{(2\pi)^3}\int d^3k (\partial_b f_0) \Omega_d = -\frac{1}{(2\pi)^3}\int d^3k (\partial_b \Omega_d) f_0.
\end{equation}
Here $f_0$ is the equilibrium Fermi distribution function in the absence of $\vec {\cal E}$. 

A non-zero antisymmetric component, ${\cal D}^{-}=({\cal D}-{\cal D}^T)/2$, of the ${\cal D}_{bd}$ tensor is a signature of a polar metal and the direction of the vector $\vec d$ with components $d_a = \varepsilon_{abc} {\cal D}_{bc}^{-}/2$ indicates the orientation of the polar axis. Thus by extracting ${\cal D}^{-}$ from measurement of the conductivity $\chi_{abc}$ tensor, important insights into the polarity of a metal can be obtained. 

The proposal of NHE resulting from the Berry curvature dipole, has recently been verified experimentally for bilayer and multilayer WTe$_2$ \cite{Ma2019,Kang2019} using Hall measurements, which show the generation of a Hall voltage $V_{yxx}^{2\omega}$ along the $y$ direction with twice the frequency $\omega$ of the applied current $I^\omega_x$ along $x$.
A Berry curvature dipole ${\cal D}_{xy}=-{\cal D}_{yx}$ and a NHE have also been predicted in LiOsO$_3$ using DFT \cite{Xiao2020}, and are awaiting experimental verification. More generally, we propose that the NHE can be used as a tool for detecting the polar metal phase in future studies.  

\subsection{Current-Induced Magnetism}

An additional non-equilibrium effect that occurs in nonmagnetic gyrotropic metals, and therefore is universal to nonmagnetic polar metals, is the kinetic magnetoelectric (KME) effect \cite{Levitov1985,Yoda2015,Zhong2016,Tsirkin2018}, The KME describes the magnetization ${\cal M}_j$ generated by an electric field $E_i$, ${\cal M}_j={\cal K}_{ij}E_i$, with the linear KME response $K_{ij}$ caused by the flow of current, in contrast to the conventional magnetoelectric effect in noncentrosymmetric magnetic insulators. 

The mechanism of the KME is very similar to that of the NHE. A net magnetization is developed in the current-induced non-equilibrium electron distribution 
corresponding to the shifted Fermi surface. The shift in the Fermi surface $\vec a_s$ is given by $-\frac{e\tau}{\hbar}E_i \hat i =-a_0 E_i \hat i$ as indicated in \textbf{Figure \ref{fig8}}\textit{c} 
and within the relaxation time approximation, the resulting magnetization is given by,  
\begin{eqnarray}
  {\cal M}_j = -\frac{a_0}{(2\pi)^3} \int d^3k~ m_j (\vec k) \partial_{k_i} \varepsilon_k \Big( \frac{\partial f_0}{\partial \varepsilon_k } \Big) E_i.
\end{eqnarray}
While the concept of KME was introduced in 1985 
\cite{Levitov1985} and later revived in the context of topological WSMs \cite{Yoda2015,Zhong2016} and 2D van der Waals metals \cite{He2020,Bhowal2020}, to the best of our knowledge the effect has not yet been studied in polar metals, even though they may have some additional functionalities. 
For example, the reversal of electric polarization in a FE metal would  switch the KME response, and the structural transition in FE-like metals would allow the effect to be switched on and off as in the case of the NHE.
From the theory perspective, it would be interesting to quantify the relationship between the response ${\cal K}_{ij}$ and the broken inversion symmetry of a polar metal. In particular, the recent prediction of $k$-space magnetoelectric multipoles \cite{BhowalCollinsSpaldin2022} in broken-inversion-symmetric systems would provide a link, in the same way that real-space ME multipoles are responsible for the conventional ME effect \cite{Spaldin2008,Spaldin2013}. Establishment of a connection between $k$-space magnetoelectric multipoles and the properties of polar metals could provide a powerful quantitative framework for properties prediction and is a promising direction for future research.

\subsection{FE Superconductivity}
A Rashba-type interaction has also been implicated as a crucial ingredient in ``FE superconductors" such as SrTiO$_3$ (STO) \cite{Kanasugi2018}. STO is a quantum PE, in which quantum fluctuations suppress the long range FE order, prohibiting ferroelectricity even at very low temperature. However, ferroelectricity can be induced through isovalent Ca doping (see the dashed blue curve in \textbf{Figure \ref{fig8}}\textit{d}), stress, or substitution of isotope $^{16}$O with $^{18}$O. In addition, STO becomes metallic on carrier doping (see the dashed brown curve in \textbf{Figure \ref{fig8}}\textit{d}) by substituting Sr with La, Ti with Nb, or by introducing oxygen vacancies, and shows non-BCS superconductivity at low temperature. 

There has been a recent surge in FE superconductor research in STO, driven by experimental evidence for a FE quantum critical point and theoretical prediction of its role in mediating superconductivity \cite{Rowley_et_al:2014,Edge2015}. The coexistence of ferroelectricity and superconductivity is now well established \cite{Rischau2017,SalmaniRezaie/Ahadi/Stemmer:2020}, and tuning of the position of the quantum quantum critical point using Ca substitution \cite{Rischau2017}, oxygen isotope substitution \cite{Stucky_et_al:2016} or strain \cite{Ahadi_et_al:2019,Herrera_et_al:2019}, have been shown to affect the superconducting $T_c$ (\textbf{Figure \ref{fig8}}\textit{d}) consistently with the predictions.  
Superconductivity has also been reported at the surfaces of quantum PE KTaO$_3$ (KTO) (since KTO is not easily chemically doped, the surfaces are required for introduction of carriers by field effects \cite{Ueno_et_al:2011} or electrostatic doping \cite{Liu2021}), again with $T_c$ correlating with proximity to FE quantum criticality.

We expect that research on polar and FE metals will increasingly extend to encompass polar and FE superconductors over the next years. Many open questions remain in the class of quantum paraelectric materials mentioned here, particularly the role of the Rashba interaction \cite{Kanasugi2018,Gastiasoro2020}, and the detailed nature and significance of the FE quantum fluctuations \cite{Enderlein2020}. In addition, extension to other polar metal classes such as FE Dirac materials \cite{Kozii2019} will undoubtedly reveal new physics and raise new questions.

\section{SUMMARY AND OUTLOOK} 
The recent discoveries of practical materials that combine the apparently contra-indicated properties of polarization and metallicity, more than half a century after the combination was first proposed, have generated a surge of activity in the field. The new materials open up a platform for investigating intriguing physical and chemical properties that emerge from such a coexistence of seemingly mutually exclusive properties, and many interesting aspects have unfolded. As with other  contra-indicated multifunctional materials, research in the polar metals field is full of challenges and many open questions need consideration from both theoretical and experimental perspective. Here we summarize the key messages of this review and point out some of the open questions that we find most intriguing for future work.  

Regarding design principles for identifying new polar metals, a common theme that emerges is the decoupled electron mechanism, in which polarity and metallicity have different chemical or spatial origins and do not interact strongly. For example, in ABO$_3$-type perovskite polar metals, polar displacement of the A-site cation can originate from geometry reasons such as BO$_6$ octahedral rotation or small size, whereas the metallic electrons occupy primarily the transition metal-$d$ states at the B site. In this context, lone-pair-active perovskite FEs, in which the stereochemically active lone pairs on A-site cations such as Pb$^{2+}$ or Bi$^{3+}$ are responsible for the FE distortion, are good candidates for further study. In non-decoupled systems, such as the conventional proper ABO$_3$-type FEs in which displacement of the B-site ion drives the ferroelectricity, the polar distortion has been shown to persist even up to a certain concentration of added itinerant charge carriers. This doping tolerance is even more robust in hyperferroelectrics and we expect similar robustness in improper FEs such as the hexagonal manganites \cite{Sai2009, Nordlander_et_al:2019}, which should be a subject of future work. Building on this newly acquired knowledge of successful design principles, searches of databases such as the Inorganic Crystal Structure Database (ICSD) \cite{BenedekBirol2016,Bennett2020}, ab-initio high-throughput structure screening \cite{Fang2020}, and machine learning predictions could further accelerate the search for new polar metals, in particular in shortlisting promising candidate materials for experimental verification. 
Finally, thin film multilayer heterostructures or superlattices offer a huge phase space of potential materials chemistries and structural modifications, in particular through coherent heteroepitaxial strain, which can modify bond lengths and octahedral rotations \cite{rondinelli2012}, templating of octahedral rotations across interfaces and electrostatic engineering \cite{Mundy_et_al:2020}.

Another direction that needs further attention is the switchability of the electric polarization, which has not been achieved in most cases or, in the case of WTe$_2$, is not  well understood, as discussed before. In this context, the recent proposal \cite{Zabalo2021} of controlling the polarization via strain gradient (flexocoupling) instead of electric field in polar metals might be a promising direction for future exploration. Achieving spin-spiral-driven polar metals as mentioned in Section~\ref{spinspiral} would also be useful as the polarization could be switched by changing the spin structure using an external magnetic field.   

Regarding properties, most predicted or observed properties of polar metals to date have been material-specific, and general properties that are features of the entire polar metal class remain to be established. For example, the link between polarity and topology in metals should be clarified. While TSMs fit the general decoupled electron argument with the localized charges at the Weyl points leading to low spectral density at the Fermi level, and many individual polar metals are predicted to be TSMs, there is no coherent understanding regarding if or when polar metals tend to become topological. Such understanding would enable the rational design of polar topological metals, and in turn  the manipulation of Weyl nodes by controlling the polar displacements. 
Conversely, we showed in this review that the non-linear Hall response and the linear KME response must occur in all polar metals by symmetry. For both of these effects, however, the number of experimental efforts to measure them, or indeed theoretical predictions of their magnitude, is scarce. Such studies would be valuable in establishing these effects as an operative method for the detection of FE-like phase transition in polar metals.

We hope that the present review will motivate future research, both theoretical and experimental, in the development of new polar metals, in the establishment of the phenomena discussed here, and in the identification of new physics that is not yet envisaged, in the coming years.

\section*{DISCLOSURE STATEMENT}
The authors are not aware of any affiliations, memberships, funding, or financial holdings that
might be perceived as affecting the objectivity of this review. 

\section*{ACKNOWLEDGMENTS}
NAS and SB were supported by the ERC under the EU’s Horizon 2020 Research and Innovation Programme grant No 810451 and by the ETH Zurich.

\bibliographystyle{ar-style3.bst}
\bibliography{Reference }

\end{document}